\documentclass[journal = jctcce, manuscript = article, layout = traditional]{achemso}
\usepackage{bm}
\usepackage{amsmath}
\usepackage{multirow}
\usepackage{graphicx}
\usepackage{color}
\usepackage{rotating}
\usepackage{booktabs,threeparttable}
\usepackage{braket}
\def \figures{.}

\def \<{\langle}
\def \>{\rangle}
\def \4sp{\ \ \ \ }
\def \etal{\textit{et al.}}

\def \anoxdz#1{ANO-RCC-{#1}ZP+$d$}
\def \dz{double-$\zeta$}
\def \tz{triple-$\zeta$}
\def \qz{quadruple-$\zeta$}
\def \eps#1{\epsilon_{#1}}

\def \cth{{\bar{H}}}       

\title{Correct quantum chemistry in a minimal basis from effective Hamiltonians}

\author{Thomas J. Watson Jr.}
\email{tjwatson@princeton.edu}
\affiliation{Department of Chemistry,
             Princeton University,
             Princeton, NJ, 08544}

\author{Garnet Kin-Lic Chan}
\email{gkchan@princeton.edu}
\affiliation{Department of Chemistry,
             Princeton University,
             Princeton, NJ, 08544}

\date{today}
\begin{document}
%
%
%
%
\begin{abstract}

We describe how to create ab-initio effective Hamiltonians that qualitatively
describe correct chemistry even when used with a minimal basis. The Hamiltonians
are obtained by folding correlation down from a large parent basis into a 
small, or minimal, target
basis, using the machinery of canonical transformations. We demonstrate the
quality of these effective Hamiltonians to correctly capture a wide range of excited states
in water, nitrogen, and ethylene, and to describe ground and excited state
bond-breaking in nitrogen and the chromium dimer, all in small or minimal basis sets.

\end{abstract}

\section{Introduction}
\label{sec:introduction}

The rapid evolution of quantum chemistry over the last decades means that in many molecules 
\cite{polyansky2003high,karton2006w4,tajti2004heat,sharma2014spectroscopic}
and even in some condensed phase systems \cite{yang2014ab}, the combination
of many-electron correlation methods with large basis sets provides predictions to 
beyond chemical accuracy of 1 kJ/mol. Despite these numerical advances,  qualitative chemical and physical reasoning necessarily
remains rooted in simple concepts.

One way to connect quantitative calculations to qualitative understanding is to construct an {effective} Hamiltonian to describe the correct correlated behaviour in terms of only 
the minimal chemical degrees of freedom, i.e. a minimal basis.
Semi-empirical methods define such Hamiltonians by fitting to precomputed observables, but 
relying on empirical parametrization
removes many advantages of predictive computation. A more satisfying route to effective minimal basis Hamiltonians is via rigorous ab-initio many-body theory. In this work,
we will construct minimal basis effective Hamiltonians by rigorous many-body canonical transformations. 
While in principle an exact procedure, in practice approximations are necessary.
We will thus be primarily concerned with addressing two questions of
approximation. First, how do we  define a simple, cheap, and stable, approximate canonical transformation to obtain the effective 
minimal basis Hamiltonian? And second, how well do such minimal basis Hamiltonians capture non-trivial chemistry, at least at a qualitative level?


We must use an {\it effective} as opposed to a {\it bare} Hamiltonian in a minimal basis 
because it is well established that quantum chemistry with the bare minimal basis 
 Hamiltonian is exceedingly poor.
This is because the electrons in filled orbitals cannot avoid each other, and the Coulomb 
interaction is felt too strongly. 
Modifying the Coulomb interaction to take into account  excursions of electrons into orbitals external to the minimal basis is referred
to as {\it folding in} the (effects of the) external orbitals. Alternatively, since the effective Coulomb interaction is decreased in magnitude,
this process is often referred to as {\it screening}.

The effective Hamiltonian of interest depends in part on the choice of many-body formalism. 
For example, in Green's function approaches, we can
define an effective two-particle (four-point) interaction operator $\Gamma(1234)$ (where the labels $1,2,3,4$ include both time and orbital indices) that yields the appropriate 
two-particle (four-point) Green's function $K(1234)$ in the minimal basis, using
a Dyson-like equation~\cite{blaizot1986quantum}
\begin{align}
K=(GG)+(GG)\Gamma (GG)
\end{align}
where $G$ is the single-particle Green's function. The interaction operator $\Gamma$ depends on time. When limited to the particle-hole channel, it is called
 the screened interaction, and is commonly computed
within the random-phase approximation~\cite{onida2002electronic,werner2010dynamical}.

Here, however, we are concerned with the effective Hamiltonian $\bar{H}$ to be used 
{\it in the many-body Schr\"odinger equation in the minimal basis},
\begin{align}
\bar{H} |\Phi_i\rangle = E_i |\Phi_i\rangle
\end{align}
The state $|\Phi_i\rangle$ exists only in the Hilbert space of the minimal basis, but is related to the exact eigenstate $|\Psi_i\rangle$ in the full space by a
many-body canonical transformation $|\Psi_i\rangle = U |\Phi_i\rangle$,  thus $\bar{H} = U^\dag H U$. Unlike
the interaction operator $\Gamma$, the effective Hamiltonian $\bar{H}$ is time-independent, but in general contains three-body and
higher body terms. The Hilbert space that $|\Phi_i\rangle$ lives in can be thought of as 
spanned  by a 
basis of quasi-particles associated with second-quantized operators $\bar{a}^{(\dag)} = U^\dag a^{(\dag)} U$.

The basic formalism of  (canonically transformed) effective Hamiltonians is very 
old and well-known. There are two families of approximation methods.
The first focuses on the effective Hamiltonian itself and dates all the way back to Van Vleck~\cite{kemble1937fundamental} and other early workers such as Brandow~\cite{brandow1967linked}, Westhaus~\cite{westhaus1973cluster}, Freed~\cite{freed1974theoretical,iwata1976nonclassical}, and others~\cite{durand2009effective}.
Also in the first family are the renormalization group approaches to $\bar{H}$,
based on successive iterative approximations, as developed by
Wegner~\cite{wegner1994flow}, Glazek and Wilson~\cite{glazek1993renormalization}, and 
White~\cite{white2002numerical}.
The second family of methods focuses more on the eigenstates $\ket{\Psi_i}$ and their
associated wave operator. 
This includes the many variants of coupled cluster theory~\cite{bartlett2007coupled}, and especially the equation-of-motion~\cite{stanton1993equation} and multireference extensions~\cite{lyakh2011multireference}.
There is much overlap between the families and there are methods which belong to both
 (such as the earlier canonical transformation work of Yanai, Neuscamman, and Chan~\cite{yanai2006canonical,yanai2007canonical,neuscamman2010strongly,neuscamman2010review},
the anti-Hermitian contracted Schr\"odinger equation of Mazziotti~\cite{mazziotti2006anti,mazziotti2007anti}, and the recent similarity renormalization group work of Evangelista~\cite{evangelista2014driven}).
However, a defining difference is that in the presence of degeneracies and strong 
correlations, the 
first family usually adopts
a ``perturb and diagonalize'' strategy, while the second
performs ``diagonalize and perturb''. This difference is one of philosophy  but can lead to 
different choices in approximations.

We are concerned here with techniques in the first family to construct effective 
minimal basis Hamiltonians. 
The earlier works by Freed, and by White, and especially the recent work by 
Yanai and Shiozaki~\cite{yanai2012canonical} are, in our view, conceptually the most 
closely related. We will explore
the approach of Yanai and Shiozaki in parallel to the new approaches in this work.
In addition, the numerical approximations we use  build on the earlier work on 
approximate canonical transformations by Yanai, Neuscamman, and Chan. 


In Section~\ref{sec:theory} we begin by precisely defining the effective Hamiltonian in a minimal basis. We then outline the earlier approximate canonical
transformation formalism of Yanai, Neuscamman, and Chan, and next the detailed
steps and approximations to construct the effective minimal basis Hamiltonians in this work.
We proceed to assess the performance of our effective Hamiltonians for a variety of 
chemical phenomena, including electronic 
excited states and  potential energy surfaces. 
We finish with a discussion of some future directions of this approach.

\section{Theory}

\label{sec:theory}
\subsection{The effective Hamiltonian}

We begin with some notation. The effective Hamiltonian folds the effects of electron
correlation from an initial large (possibly infinite) ``parent'' basis down
to a smaller ``target'' basis. We label orbitals in the parent basis (assumed orthogonalized)  by indices $\{\kappa, \lambda, \mu, \nu\}$. Thus the
parent basis Hamiltonian  is written as
\begin{equation}
   {H} = \sum_{\mu \nu} h_{\mu \nu} {E}_{\mu\nu}
           + \frac{1}{4} \sum_{\mu \nu \kappa \lambda} V_{\mu \nu \kappa \lambda} {E}_{\mu \nu \lambda \kappa}
\end{equation}
where we use the spin-summed excitation operators ${E}_{\mu\nu}=\sum_{\sigma \in \{\alpha,\beta\}} a^\dag_{\mu \sigma} a_{\nu\sigma}$, and $V_{\mu \nu \kappa \lambda}$ represents the anti-symmetrized two-electron integral operator.

We label the target basis (assumed orthogonalized) by  $\{p, q, r, s\}$. 
In this work, we take the target basis to be spanned by a
set of  minimal Gaussian basis functions, although in principle any small basis can be used.
The (orthogonal) functions that are in the parent basis {\it but which  live outside
of the target basis}, define the external space; we will label these by $\{ x, y\}$. 
When the parent basis is formally infinite (as in R12/F12 theory), the external space
is represented, as necessary, by its complementary auxiliary basis~\cite{valeev2004improving} 
(discussed in detail in Section~\ref{sec:F12}). Note that for two arbitrary 
finite parent and target Gaussian bases,
the target basis is not usually a subspace of the larger parent basis. In that case,
we consider the parent basis to be the {\it union} of the target basis and
the original parent basis. 
For example, if using an ANO-RCC-MIN target basis~\cite{roos2005new} and 
an aug-cc-pVQZ~\cite{dunning1989basis} parent basis, we take the parent basis in
numerical calculations to be ANO-RCC-MIN+aug-cc-PVQZ. The
 dimension of the external space is then the same as that of the original parent basis.

The exact effective Hamiltonian in the target basis is
\begin{equation}
   \cth = \bar{h}_{pq} {E}_{pq} + \frac{1}{4} \bar{V}_{pqrs} {E}_{pqrs} + 
\frac{1}{36} \bar{W}_{pqrstu} {E}_{pqrstu} + \ldots
\label{eq:cth}
\end{equation}
where $\ldots$ indicates additional higher-body interactions.
$\cth$ is related to $H$ by the canonical transformation operator $e^A$, using
the Baker-Campbell-Hausdorff (BCH) expansion,
\begin{align}
\cth & = e^{A^\dag} H e^A \notag\\
&={H}
              + \left[{H}, {A} \right]
              + \frac{1}{2}\left[[{H},{A}],{A}\right] + \ldots \label{eq:bch}
\end{align}
where $A$ is the antihermitian excitation and de-excitation operator between the target and external space. $A$ can be written in terms of 1-body, 2-body, and higher components,
\begin{align}
   {A} &= {A}_1 + {A}_2 + \ldots \notag \\
           &= \sum_{px} t_{px} \left({E}_{xp} - {E}_{px} \right)
              + \frac{1}{4}\sum_{ pqxy} t_{pqxy} \left({E}_{xypq} - {E}_{pqxy} \right) + \ldots \label{eq:amps}
\end{align}

The amplitudes of $A$: $t_{px}$, $t_{pqxy}$, etc., are chosen to make all matrix elements 
between the target and external spaces vanish. 
This leads to  equations of the form of generalized Brillouin conditions~\cite{kutzelnigg1979generalized,yanai2006canonical},
\begin{align}
\langle \Phi| [\cth, \left({E}_{px} - {E}_{xp} \right)] | \Phi\rangle &= 0, \nonumber\\
\langle \Phi| [\cth, \left({E}_{pqxy} - {E}_{xypq} \right)] | \Phi\rangle &= 0, \ldots \label{eq:brill}
\end{align}
where $\ket{\Phi}$ is {\it any} state in the target space. The complete Brillouin conditions
define $\cth$ uniquely, up to decoupled rotations within the target and external spaces separately.

In the above equations we immediately observe
the need for approximation. This arises in (i) truncating $A$ in Eq. (\ref{eq:amps}), (ii) handling the BCH expansion of $\cth$ and the many-body interactions in Eqs. (\ref{eq:cth}), (\ref{eq:bch}), and (iii) solving the amplitude equations, Eq. (\ref{eq:brill}). We now 
discuss approximations in each of these areas.

\subsection{Approximate canonical transformations}
 

In defining approximate canonical transformations we are motivated by the
prior experience of Yanai, Neuscamman, and Chan~\cite{yanai2006canonical,yanai2007canonical,neuscamman2009quadratic,neuscamman2010review} and Yanai and Shiozaki\cite{yanai2012canonical}, 
and by the  accuracy requirements in this work.
Our goal here is  {\it qualitative} accuracy for a broad range of phenomena, rather than highly quantitative (e.g. 1 kcal/mol) chemical accuracy 
for a particular single target state at a given geometry.

For (i) and (ii), we re-use the main ideas in Yanai, Neuscamman, and Chan, namely
we truncate amplitudes at the two-body level ($A=A_1+A_2$) and limit the effective Hamiltonian to two-body interactions by approximating higher-body terms. 
For qualitative accuracy, it is reasonable to truncate the BCH expansion
through second-order perturbation theory
as in the work of Yanai and Shiozaki. This gives $\bar{H}$ as
\begin{align}
\cth \approx H  + \left[{H}, {A} \right]
              + \frac{1}{2}\left[[{F},{A}],{A}\right] \label{eq:hylcth}
\end{align}
where $F$ is a Fock operator (defined more precisely below). 
If $F$ is defined using the Hartree-Fock density matrix, the expectation value 
of $\cth$ with the Hartree-Fock determinant $\bra{\Phi_D}\bar{H}\ket{\Phi_D}$ is the Hylleraas second-order energy functional.

Note that although accurate only through second-order, the effective Hamiltonian in Eq.~(\ref{eq:hylcth})  already involves {\it three}-body interactions, generated by
the first and second commutators. However, approximating $\bar{H}$ by a two-body Hamiltonian
is well supported by the success of semi-empirical and model Hamiltonians, 
which universally assume only two-body interactions. Decomposing  three- and higher-body interactions into effective two-body interactions was already considered by Iwata and Freed in the effective valence Hamiltonian theory~\cite{iwata1976nonclassical}. In the canonical transformation
theory of Yanai and Chan, this decomposition was systematized
using the generalized normal ordering of Mukherjee and Kutzelnigg~\cite{kutzelnigg1997normal} (an extension of the density cumulant expansion to operators)~\cite{mazziotti1998approximate,mazziotti1998contracted,kutzelnigg1999cumulant}. This re-expresses
the 3-body operators generated above, $E^{pqr}_{stu}$, by 
\begin{align}
a^{pqr}_{stu} \approx 9 (\gamma^p_s \wedge a^{qr}_{tu}) - 36 (\gamma^p_s \wedge \gamma^q_t \wedge \gamma^r_u) + 9 (\gamma^{pq}_{st} \wedge a^r_u) + 24 (\gamma^p_s \wedge \gamma^q_t \gamma^r_u) - 9 (\gamma^{pq}_{st} \wedge \gamma^r_u)
\label{eq:gno}
\end{align}
where we drop explicit 3-body fluctuation operators and 3-body density cumulants. 
(The above term $a^{pqr}_{stu}$ is expressed in terms of spin-orbitals. The correct spin-summed expression
is given in Ref.~\cite{shamasundar2009cumulant}, but we use the spin-orbital
decomposition in our implementation). 
Since both the cumulants and the expectation value of the 3-body fluctuation operators vanish for
a determinant state,  truncating the 3-body terms in Eq.~(\ref{eq:gno}) 
preserves the expectation value of $\cth$ with any determinant, and thus the value
of the standard Hylleraas functional.
Denoting the truncated normal ordering approximation in Eq.~(\ref{eq:gno}) by the subscript $(1,2)$, we obtain
 the approximate two-body $\bar{H}_{(1,2)}$ as
\begin{equation}
  \cth_{(1,2)} = {H}
              + \left[{H}, {A} \right]_{(1,2)}
              + \frac{1}{2}\left[[{F},{A}],{A}\right]_{(1,2)} \label{eq:cth12}
\end{equation}
When not ambiguous, we  drop the subscript $(1,2)$ in the labelling of $\cth_{(1,2)}$,
understanding that it is defined as above.

Both the Fock matrix $F$ and the normal ordering approximation in Eq.~(\ref{eq:gno})
require density matrices, thus introducing state-specific information. To
minimize the amount of initial state-specific 
computation with the bare Hamiltonian (in keeping with the perturb and diagonalize
philosophy) we use only density matrices from 
the Hartree-Fock ground-state in the target basis unless otherwise specified.

\subsection{Approximate amplitudes}

\label{sec:amps}

We next discuss how to determine the amplitudes. In the original canonical transformation
work of Yanai and Chan, the amplitude equations are solved 
with respect to an initial (multireference) state $|\Psi_0\rangle$, typically a ground-state complete active space wavefunction, using the normal-ordered approximate $\bar{H}$ and
the Brillouin conditions in Eq.~(\ref{eq:brill}) with $|\Phi\rangle = |\Psi_0\rangle$.
This provides a route to add dynamic correlation to the strong correlations presumed captured by $\ket{\Psi_0}$. 
However, there are several reasons {\it not} to use this strategy here. First, solving the amplitude equations for a multireference $\ket{\Psi_0}$ is expensive  - similar in scaling to, but
several times the cost of, a coupled cluster singles doubles calculation. This  is overkill for the qualitative accuracy we target.
More importantly,  there are fundamental numerical issues
when solving the Brillouin conditions. The Jacobian of Eq.~(\ref{eq:brill}) has small eigenvalues, and without a cutoff of these eigenvalues
hundreds of iterations of the equations may be required. These small Jacobian eigenvalues also appear in internally contracted multi-reference coupled cluster theory~\cite{banerjee1981coupled,evangelista2011orbital,hanauer2011pilot,datta2011state}
as well as in the anti-Hermitian contracted Schr\"odinger equation~\cite{mazziotti2006anti,mazziotti2007anti} (mathematically equivalent to a Schr\"odinger picture formulation of the Brillouin conditions)  
leading to  convergence difficulties in both these methods.  

Further, at a fundamental level, the small Jacobian eigenvalues arise because the 
amplitude equations are solved for a single reference state $|\Psi_0\rangle$, which may have 
no weight in certain orbitals or electron configurations in the target space. Thus the amplitudes
obtained are biased towards the reference state.
This is an advantage for high accuracy for a given state (as in the original
canonical transformation theory, or as in coupled cluster theory) but is a liability for a
qualitative effective Hamiltonian intended to describe many states on an equal footing.

For both these reasons, here we construct approximate amplitudes in a manner that 
does not require 
the iterative solution of ill-conditioned and state-specific equations. 
We start with an effective Hamiltonian defined by the Hylleraas expression in Eq.~(\ref{eq:cth12}),
and
use a Fock operator $F$ built from the Hartree-Fock density matrix in the small target basis.
Re-expressing this in the parent basis, $F$ can be written in semi-canonical form, for computational
efficiency, as
\begin{equation}
   \mathbf{F} = \left[ \begin{array}{cccc}
          F_{ij}\delta_{ij} & 0 & \ddots  \\
          0 & F_{ab} \delta_{ab} & \ddots \\
          \ddots & \ddots & F_{xy} \delta_{xy}  \\
   \end{array} \right] \label{eq:fsemi}
\end{equation}
where $i,j$ and $a,b$ denote occupied and virtual molecular orbitals of the Fock operator
in the target basis, respectively.
The solution of the Brillouin condition with $\ket{\Phi}=\ket{\Phi_D}$ (where $\Phi_D$
is the target basis Hartree-Fock Slater determinant)
is the Moller-Plesset first-order wavefunction, with amplitudes 
\begin{align}
   t_{ix} &= \frac{F_{ix}}{\eps{i} - \eps{x}} \notag\\
   t_{ijxy} &= \frac{V_{ijxy}}{\eps{i} + \eps{j} - \eps{x} - \eps{y}} \label{eq:mp2amp}
\end{align} 
where $\eps{i}=F_{ii}$, $\eps{x}=F_{xx}$, respectively.
 We can now use these amplitudes to construct $\bar{H}$.
The singles and doubles amplitudes respectively add the Hartree-Fock basis set correction 
from the external space, and the correlated MP2 contribution from the external space, to the zeroth order Hartree-Fock energy of $\bar{H}$.

Note, however, that the amplitudes in Eq.~(\ref{eq:mp2amp}) are defined only between  orbitals  occupied in $\ket{\Phi_D}$ $i,j$, and the external orbitals $x,y$.
To decouple other low-lying states in the target basis from the external space, we consider using other (non-ground-state) 
determinants $\ket{\Phi_D'}$ to  define the amplitudes in Eq.~(\ref{eq:mp2amp}), which occupy ``active virtual'' orbitals $a,b,c$ (orbitals  not occupied in the lowest determinant). This defines additional amplitudes involving the active virtuals, such as
\begin{align}
t_{ax} &= \frac{F_{ax}}{\eps{a} - \eps{x}} \notag\\
t_{abxy}&=\frac{V_{abxy}}{\eps{a} + \eps{b} - \eps{x} - \eps{y}} \label{eq:amp2}
\end{align}
and similar  amplitudes with mixed (active and occupied) indices, such as $t_{ia}^{xy}$. 
Such amplitudes involving ``active virtuals'' are   omitted
 in the definition of the effective Hamiltonian in equation-of-motion coupled cluster theory, because they annihilate the lowest Hartree-Fock ground-state $\ket{\Phi_D}$ and do not change the 
ground-state energy.
However, they are necessary to remove the bias towards the ground-state in the resulting effective
Hamiltonian.
Unfortunately, if we use the additional amplitudes as naively  written in Eq.~(\ref{eq:amp2}),
we obtain very poor results. This is because the virtual eigenvalues 
$\epsilon_a$, $\epsilon_b$ appearing in Eq.~(\ref{eq:amp2})
are  determined for the {\it ground-state} density matrix and Fock operator, 
rather than for the Fock operator corresponding to the excited state determinant $\ket{\Phi_D'}$. As the eigenvalue difference 
between the active virtual orbitals and external orbitals  often vanishes, Eq.~(\ref{eq:amp2}) can even yield divergent amplitudes. 

Physically, the energy of an electron in one of the active virtual orbitals $\phi_a$ is not well approximated by $\epsilon_a$, but is rather much closer
to the HOMO energy level. This is because in Hartree-Fock theory the virtual energy levels are 
optimized in the field of $N$ rather than $N-1$ electrons.
The appropriate orbital relaxation effect would be properly included if we retained three-body operators in the effective Hamiltonian (similar
to treating triples excitations in coupled cluster theory) but are not properly captured in the $(1,2)$ approximation. To partially take into account this 3-body effect 
we now introduce a simple approximation. In  the definition of amplitudes
involving an active virtual orbital, {\it we replace the
corresponding virtual Fock energy $\epsilon_a$ 
by a single modified orbital energy, $\bar{\epsilon}_{a}$}. In the simplest
case, we replace $\bar{\epsilon}_a$ by the HOMO energy, but we can also view $\bar{\epsilon}_a$
as an adjustable parameter. (This modification of the active virtual denominators can also be justified from the viewpoint of degenerate perturbation theory, as argued
by Iwata and Freed, who placed {\it all} occupied and active (valence) orbitals 
at the same average 
energy thus creating a truly degenerate zeroth order Hamiltonian~\cite{iwata1976nonclassical}). 
Thus, 
\begin{align}
   t_{ax} &= \frac{F_{ax}}{\eps{x} - \bar{\epsilon}_a} \notag\\
   t_{a jxy} &= \frac{V_{a jxy}}{\eps{x} + \eps{y} - \bar{\epsilon}_a - \eps{j}} \notag\\
   t_{a bxy} &= \frac{V_{a bxy}}{\eps{x} + \eps{y} - 2\bar{\epsilon}_a}
\end{align}
{With this regularization, we define $A$ to include all amplitudes involving the additional active virtual labels, except for
active-active to active-external ($abcx$).
Including the latter class of amplitudes tends to lead to significantly worse results,
presumably because it requires  a more rigorous treatment of the higher body effects than
this simple effective denominator}. 


\subsection{Comparison to the approach of Yanai and Shiozaki}

\label{sec:F12}

In their work on canonical transcorrelated Hamiltonians, Yanai and Shiozaki used a simple definition of the amplitudes that also does
not require the solution of amplitude equations, and which is appropriate when both the target 
 and parent basis are large.
Following Ten-no~\cite{ten2004explicitly}, the first-order MP2 amplitudes in an infinite external basis 
are fixed by the cusp condition. Using an F12 factor $f=\exp(-\gamma r_{12})$
to represent the beyond linear terms in the $r_{12}$ dependence, one obtains MP2 amplitudes given by
\begin{align}
t_{ij \alpha \beta}(F12) = \frac{1}{8}\<\alpha \beta|{Q}_{12}{F}_{12}|ij\>
                                                            +\frac{3}{8}\<\alpha \beta|{Q}_{12}{F}_{12}|ji\> \label{eq:F12amps}
\end{align}
where $\alpha, \beta$  label the formal infinite external parent basis, and $Q_{12}$
is the strong orthogonality projector~\cite{yanai2012canonical}.

In this work, we replace the formal infinite basis by the large parent basis and  view the 
formula (\ref{eq:F12amps}) as a way to provide the corresponding amplitudes $t_{ix}, t_{ax}, t_{ijxy}$ etc. 
We can include orbital relaxation effects by defining the singles as in Eq.~(\ref{eq:mp2amp}).
$Q_{12}$ ensures that the F12 excitations are orthogonal to those in the target basis since
\begin{align}
Q_{12} =  (1-{p(1)})(1-{p(2)}) -  x(1)x(2)
\end{align}
and $p(1)$ denotes a (one-particle) projector onto the orbitals (occupied
and active virtual) in the target space, and $x(1)$, to the external orbitals. 

Using these F12-derived amplitudes, we then construct $\cth$ directly 
using the BCH expansion in Eq.~(\ref{eq:hylcth}). Alternatively, Yanai and Shiozaki 
used approximation ``C'' of K{e}d\v{z}uch {\etal }~\cite{kedvzuch2005alternative} to compute the double commutator $[[F,A],A]$:
we denote the corresponding approximation, F12(C). The effective Hamiltonians
computed in either approach are identical in the limit of a large parent basis, but for rapid convergence,  one should choose the parent basis to be an auxiliary basis set 
specifically constructed for use in F12 theories.

Note that Eq.~(\ref{eq:F12amps}) defines not only
the standard occupied to external amplitudes $t_{ij}^{xy}$, 
but also active virtual to external excitations $t_{ab}^{xy}$ as well.
Thus using the complete set of F12 amplitudes (as in the earlier canonical transcorrelated Hamiltonian of Yanai and Shiozaki) also dresses the Hamiltonian for excited states.
(Note that, following Yanai and Shiozaki, we do not include active-active to active-external
semi-internal excitations).
 However, the amplitudes target only the universal part of the short-range correlations, 
as defined by the Coulomb cusp, and thus retain no state-specific information. Such state 
information is  only  represented by amplitudes in the target space.
For
this reason, we expect effective Hamiltonians derived using the universal F12 amplitudes 
to be less accurate than those obtained using the orbital MP2 coefficients in section 
\ref{sec:amps} when folding into small target bases, such as a minimal basis.

\section{Results and discussion}

\subsection{Effective minimal basis Hamiltonian energies}


As a first check of our effective Hamiltonian construction, we 
compute second order perturbation theory (MP2) ground-state energies 
 in the parent basis, using the bare Hamiltonian $H$,
and the corresponding MP2 ground-state energies in the target basis, using the
effective Hamiltonian defined in Eq.~(\ref{eq:cth12}). For comparison,
we also compute the MP2 ground-state energies in the target basis with the bare Hamiltonian,
to show the effects of folding.
To construct $\bar{H}$ we use both explicit orbital amplitudes (Sec.~\ref{sec:amps}) as well as
F12 amplitudes (Sec.~\ref{sec:F12}).
We recall that despite the $(1,2)$ approximation in the commutator expansion
for $\bar{H}$, the MP2 energies (using explicit orbital amplitudes) 
in the parent and target basis should match exactly, up to
orbital relaxation effects. The orbital relaxation effects
are not captured completely simply because they are treated to infinite order in the
Hartree-Fock calculation in the parent basis, but  only to second order
in our effective Hamiltonian.

Tables~\ref{tab:mp2foldh2o}, \ref{tab:mp2foldn2} give the errors
in the effective Hamiltonian MP2 energies relative to the parent basis
for the water and nitrogen molecules. 
We use the single-$\zeta$ contraction of Roos' ANO family of bases sets, labeled ANO-RCC-MIN, 
and Dunning's family of cc-pVXZ (X $=$ D, T), labeled DZ and TZ target bases, and an 
aQZ parent basis
 (aXZ is Dunning's aug-cc-pVXZ basis~\cite{kendall1992electron}). 
{The quoted reference MP2 aQZ energy does not include
the additional basis functions from the target basis, but this has a
negligible effect relative to the errors that we are discussing; for example,
the MP2 water energy using the aQZ basis is -76.38278 $E_h$, while using
the union of the MIN and aQZ basis, it is {-76.38405} $E_h$.}

For $\bar{H}$ constructed from orbital amplitudes, the error in 
the effective ground-state energy, due to 
incomplete orbital relaxation, is small: {[24]} m$E_h$ and {[37]} m$E_h$ for water
and nitrogen respectively, even in the smallest ANO-RCC-MIN target basis.
This compares quite favourably with the error in the bare Hamiltonian ANO-RCC-MIN MP2 energy: 
{412} m$E_h$ and {529} m$E_h$ for
water and nitrogen respectively. The error from incomplete orbital relaxation
decreases as the target basis size increases.

\begin{table}
\caption{ Difference between the MP2 ground-state energy
in various target bases (ANO-RCC-MIN, DZ, TZ),
and the MP2 aQZ (parent basis) ground-state
energy, for the water molecule at the equilibrium geometry {$R={1.80847~\text{\AA}}, \theta={104.5^{\circ}}$}. 
The $H$ column denotes the (standard) MP2 calculation with the normal
 bare Hamiltonian. The other columns refer to the calculations
using the effective Hamiltonian.
Orb. denotes $\bar{H}$ using orbital amplitudes, F12 denotes  F12 amplitudes, F12(C) denotes using approximation ``C'' in the double commutator, and F12+$A_1$ denotes
additional singles excitations. All energies in $E_h$.
The MP2 aQZ energy is {$-76.38278$ $E_h$}. (The union of ANO-RCC-MIN and aQZ MP2 energy is $-76.38405$ $E_h$).
\label{tab:mp2foldh2o} }
\begin{tabular}{ccccccccc}
\hline
\hline      
Target Basis &  Error ($H$) & \multicolumn{4}{c}{Error ($\bar{H}$)} \\
      &               & Orb. & F12  & F12(C) & F12+$A_1$\\
\hline
\hline
ANO-RCC-MIN & 0.4129 &  0.0247 & 0.2374 & 0.1933 & 0.1203 \\
DZ       & 0.1520 &  0.0330 & 0.0760 & 0.0930 & 0.0412 \\
TZ       & 0.0505 &  0.0095 & 0.0197 &  -0.0337 & 0.0119\\
\hline
\hline
\end{tabular}
\end{table}

\begin{table}
\caption{ Difference between the MP2 ground-state energy
in various target bases (ANO-RCC-MIN, DZ, TZ),
and the MP2 aQZ (parent basis) ground-state
energy, for the nitrogen molecule at the equilibrium geometry {$R=1.09768\text{\AA}$}.
The $H$ column denotes the (standard) MP2 calculation with the normal
 bare Hamiltonian. The other columns refer to the calculations
using the effective Hamiltonian.
Orb. denotes $\bar{H}$ using orbital amplitudes, F12 denotes  F12 amplitudes, F12(C) denotes using approximation ``C'' in the double commutator, and F12+$A_1$ denotes
additional singles excitations. All energies in $E_h$.
The MP2 aQZ energy is {$-109.45124$ $E_h$}. (The union of ANO-RCC-MIN and aQZ MP2 energy is -109.45259 $E_h$). 
\label{tab:mp2foldn2} }
\begin{tabular}{ccccccccc}
\hline
\hline      
Target Basis &  Error ($H$) & \multicolumn{4}{c}{Error ($\bar{H}$)} \\
      &               & Orb. & F12  & F12(C) & F12+$A_1$\\
\hline
\hline
ANO-RCC-MIN & 0.5291 &  0.0374 & 0.3396 & 0.2531 & 0.1249 \\
DZ       & 0.1865 &  0.0324 & 0.0787 & 0.0157 & 0.0453 \\
TZ       & 0.0683 &  0.0075 & 0.0184 &  -0.0364 & 0.0107\\
\hline
\hline
\end{tabular}
\end{table}

For $\bar{H}$ constructed from the  F12 amplitudes, using an aQZ CABS basis, the
errors are somewhat larger. (These errors are measured relative to the MP2 aQZ energy. We could measure the error relative to
the MP2-F12 aQZ energy, but as the difference between the MP2 and MP2-F12 aQZ energies is 
small on the scale of errors we are discussing ($1.3$ m$E_h$ for the water molecule) this
would not change the conclusions. A measure of the CABS basis completeness is given by the
difference between the F12 and F12(C) columns.)
We observe several important things. First, 
 the difference between the F12 and F12+$A_1$ columns, shows that
the CABS single particle amplitudes are very important, as they
capture orbital relaxation. Second, 
the F12 amplitudes lead to a significantly less accurate $\bar{H}$  than
 the explicit MP2 orbital amplitudes.
This is because the F12 amplitudes do not capture the non-universal
part of the short-range correlation. 

\subsection{Excitation energies}

The purpose of the effective Hamiltonian, is, of course, not simply to reproduce the
ground-state calculation from which it is constructed, but to be able
to use it in new calculations. We therefore now examine
the accuracy of the effective Hamiltonians for
the {\it excitation} energies of water, nitrogen, and ethylene. 
Density matrix renormalization group (DMRG)
was used for the water and nitrogen effective Hamiltonian and parent bases 
excited state calculations. The DMRG calculations used up to $M$=4000
(with $S^2$ symmetry) for both molecules and all states,
and are converged to microHartree level ($1s$ electrons
in nitrogen were kept frozen for all calculations). Ethylene effective
Hamiltonian and parent bases excitation energies were computed 
using the equation of motion coupled cluster
with connected single and double excitations and a perturbative treatment of triples (EOM-CCSD(T)).


Tables~\ref{tab:h2oexcite} and \ref{tab:n2excite} give the lowest few excitation energies for  
the water and nitrogen molecules using the aforementioned ANO-RCC-MIN target basis, and Dunning's DZ parent basis. 
We first examine the water molecule. The excitation energies using the bare ANO-RCC-MIN Hamiltonian 
are very poor, with a maximum error of 5.40 eV. The effective Hamiltonian 
using all orbital amplitudes (shifting the active virtual energies to $\bar{\epsilon}_a=0.2$ $E_h$)
yields a very significant improvement with a maximum error now of only 0.45 eV, quite
surprising for a calculation in a minimal basis!
For comparison, if we use F12 amplitudes the errors
of the effective Hamiltonian  are essentially unchanged
with respect to the bare Hamiltonian $H$, with a maximum error of -5.43 eV. 
This shows that the 
F12 amplitudes do not properly capture differential correlation between the
ground and excited state in this molecule. Some part of the
differential correlation is recovered using the additional $A_1$ amplitudes (F12+$A_1$),
reducing the maximum error to 0.91 eV, but this is still worse than the effective Hamiltonian
using explicit orbital amplitudes.

An important column in the table is the one labelled ``no act.''. These
show  results where ``active''
orbitals are omitted from the amplitudes in the $\bar{H}$ construction. This
has no effect on the ground-state energies in the previous section, but as we argued
will significantly affect the excitations. We see
that this effective Hamiltonian - which is in essence the Hamiltonian used
in equation-of-motion coupled cluster theories - yields even {\it worse} excitation energies
than the bare Hamiltonian, since it overcorrelates the ground-state and leads to unbalanced
treatment of excitations. In equation-of-motion coupled cluster, this imbalance is ameliorated
by rediagonalizing in the full orbital space (including
the external orbitals), not just a small ``active'' space as
is implicitly done here by diagonalizing in the minimal basis.

Column ``$[[H, A], A]$'' shows the results
of using the full Hamiltonian operator, rather than the Fock operator, in the double commutator
contribution to $\bar{H}$   in Eq.~(\ref{eq:cth12}). 
Interestingly we find that the results using the Fock operator are uniformly {\it better}
than using the true $H$. We attribute this to some form of error cancellation.

\begin{table}[ht!]
  \caption[Excitation energies for water.]
          {Errors of effective Hamiltonian excitation energies 
compared to parent basis excitation energies of water (computed using DMRG with
parameters mentioned in the text).
 Target (small) basis is {ANO-RCC-MIN} and parent basis is cc-pVDZ.
           Units are in electron Volts. State notation {X.Y.Z denotes {\textit{Multiplicity.Irrep.State}} (with \textit{State} being energy ordered)}. }
  \label{tab:h2oexcite}\centering
  \begin{threeparttable}
  \begin{tabular}{lcccccccc}
  \hline
        & Ref.   &  Bare $H$     & \multicolumn{6}{c}{$\cth$} \\
  \cline{4-9}
  State &  DZ & ANO-RCC-MIN & Orb.\tnote{a} & $[[H,A],A]$\tnote{a} & no act. & F12 & F12(C) & F12+$A_1$ \\
  \hline
   3.3.1 &  7.46 & 3.49 &  -0.36 & 1.34 & 6.48& 3.93 & 3.88  &  -0.91  \\
   1.3.1 &  8.13 & 4.24 &  -0.09 & 1.36 & 7.23 & 4.64 & 4.47 &  -0.69 \\
   3.2.1 &  9.73 & 4.56 &  -0.08 & 1.93 & 7.20 & 4.85 & 4.82 &  -0.67 \\
   3.1.1 &  9.89 & 3.24 &  -0.38 & 6.86 & 6.31 & 3.48 & 3.40 &  -0.86 \\
   1.2.1 & 10.15 & 5.04 & 0.03 & 1.92 & 7.62  & 5.27 & 5.14 &  -0.55 \\
   1.1.2 & 10.80 & 4.43 &  -0.23 & 1.07 & 7.66& 4.48 & 4.30  &  -0.44 \\
   3.4.1 & 11.90 & 3.59 &  -0.28 & 1.57 & 6.16 & 3.83 & 3.74&  -0.85  \\
   1.4.1 & 12.86 & 5.40 & 0.45 & 2.14 & 8.21& 5.43 & 5.21 & 0.11  \\
  \hline
  \end{tabular}
  \begin{tablenotes}
     \item[a] $\epsilon_a$ set to $-0.20$~a.u. 
  \end{tablenotes}
  \end{threeparttable}
\end{table}
%
%
%
%
\begin{table}[ht!]
  \caption[Excitation energies for nitrogen.]
          {Errors of effective Hamiltonian excitation energies 
compared to parent basis excitation energies of water ({computed using DMRG with
parameters mentioned in the text}).
 Target (small) basis is {ANO-RCC-MIN} and parent basis is cc-pVDZ.
           Units are in electron Volts. State notation {X.Y.Z denotes {\textit{Multiplicity.Irrep.State}} (with \textit{State} being energy ordered)}. }
  \label{tab:n2excite}\centering
  \begin{threeparttable}
  \begin{tabular}{lcccccccc}
  \hline
        &    &       & \multicolumn{6}{c}{$\cth$} \\
  \cline{4-9}
  State &  DZ & ANO-RCC-MIN & Orb.\tnote{a} & $[[H,A],A]$\tnote{a} & no act.  & F12 & F12(C)& F12+$A_1$\\
  \hline
   3.6.1 &  7.83 & 0.57 & 0.28 & 0.64 & 1.90  & 0.77 & 0.67 & 0.33\\
   3.3.1 &  8.12 &  -0.56 & 0.68 & 0.84 & 2.58 &  -0.47 &  -0.57& 0.56  \\
   3.5.1 &  9.13 & 0.85 & 0.42 & 0.60 & 2.18  & 0.95 & 0.75& 0.46 \\
   1.3.1 &  9.54 &  -0.66 & 0.77 & 0.93 & 2.72 &  -0.50 &  -0.79 & 0.50 \\
   3.5.2 &  9.93 & 0.71 & 0.34 & 0.48 & 2.24  & 0.76 & 0.52 & 0.29\\
   1.5.1 & 10.26 & 1.10 & 0.54 & 0.58 & 2.47& 1.12 & 0.84  & 0.57 \\
   1.6.1 & 10.65 & 1.05 & 0.51 & 0.47 & 2.46 & 1.04 & 0.69 & 0.47 \\
  \hline
  \end{tabular}
  \begin{tablenotes}
     \item[a] $\epsilon_a$ set to $-0.30$~a.u. 
  \end{tablenotes}
  \end{threeparttable}
\end{table}

The excitation energies for nitrogen in Table~\ref{tab:n2excite} follow a similar trend
to those seen for water. The main difference is that the excitation
energies from the bare Hamiltonian in the target ANO-RCC-MIN basis are 
not too far from those in the parent DZ basis (maximum error of -1.1 eV). We
thus expect  folding to an effective minimal basis Hamiltonian to yield less of an improvement,
and indeed that is the case. Using the effective Hamiltonian with orbital amplitudes,
the maximum error is reduced to -0.77 eV. The relative performance of the F12 amplitudes
and orbital amplitudes follow a similar trend to what is seen in the water molecule:
the F12 amplitudes alone lead to little or no improvement in the excitation energies. 

\begin{figure}[h!]
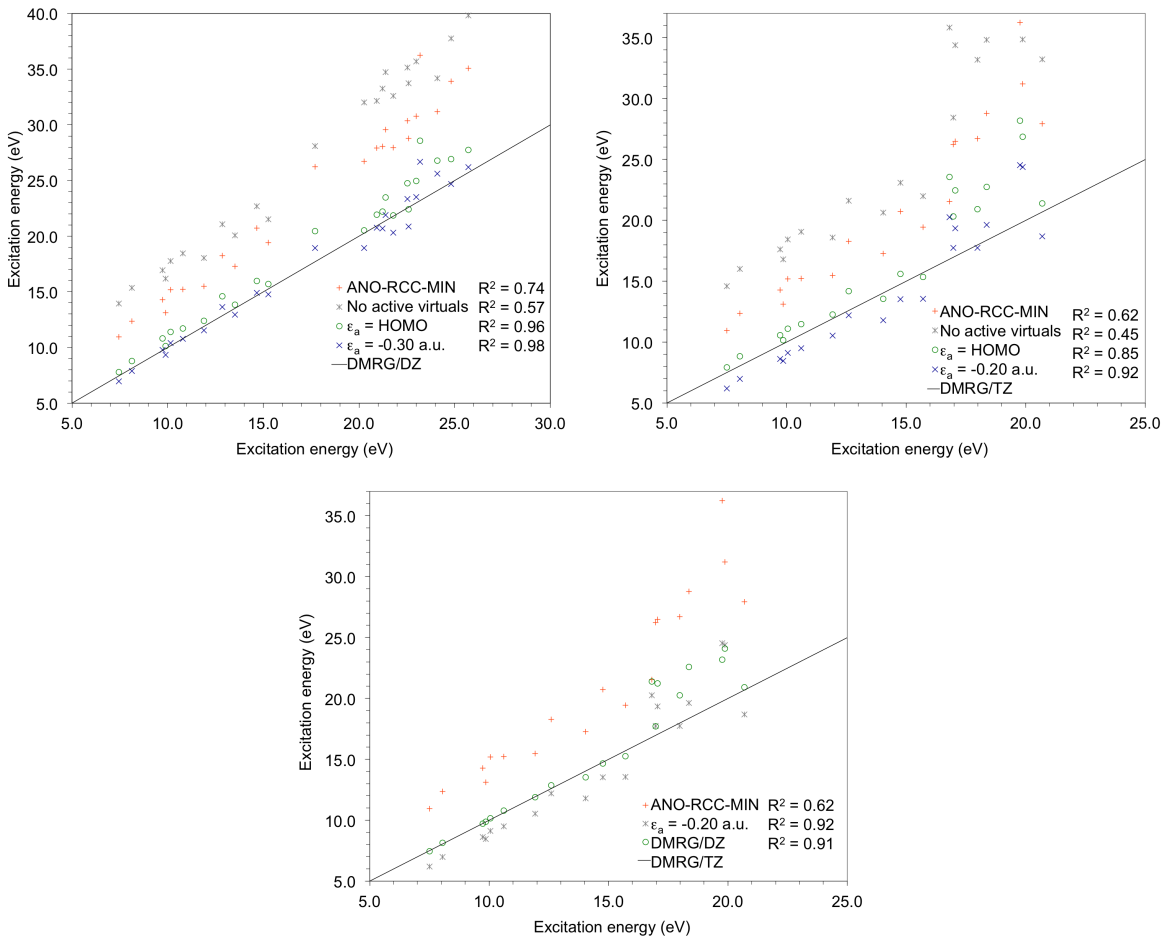

 \includegraphics[scale=0.30]{{{\figures/Water.MIN_DZ.DZ.corrplot}}}
 \includegraphics[scale=0.30]{{{\figures/Water.MIN_TZ.TZ.corrplot}}}\\
 \includegraphics[scale=0.30]{{{\figures/Water.MIN_TZ.DZ.corrplot}}}
   \caption{Correlation plots for excitation energies of the water molecule 
     with the target ANO-RCC-MIN effective Hamiltonian, compared to the excitation energies in the parent basis (cc-pVDZ (left), cc-pVTZ (right)).  The bottom chart compares
 the excitation energies of a target ANO-RCC-MIN effective Hamiltonian 
(parent cc-pVTZ basis) and excitation energies of
 a bare cc-pVDZ Hamiltonian with the parent cc-pVTZ excitation energies, to
see the ``$\zeta$'' level of improvement due to canonical transformation.  (Excitation energies computed using DMRG).
\label{fig:water.spectrum}}
\end{figure}
\begin{figure}[h!]
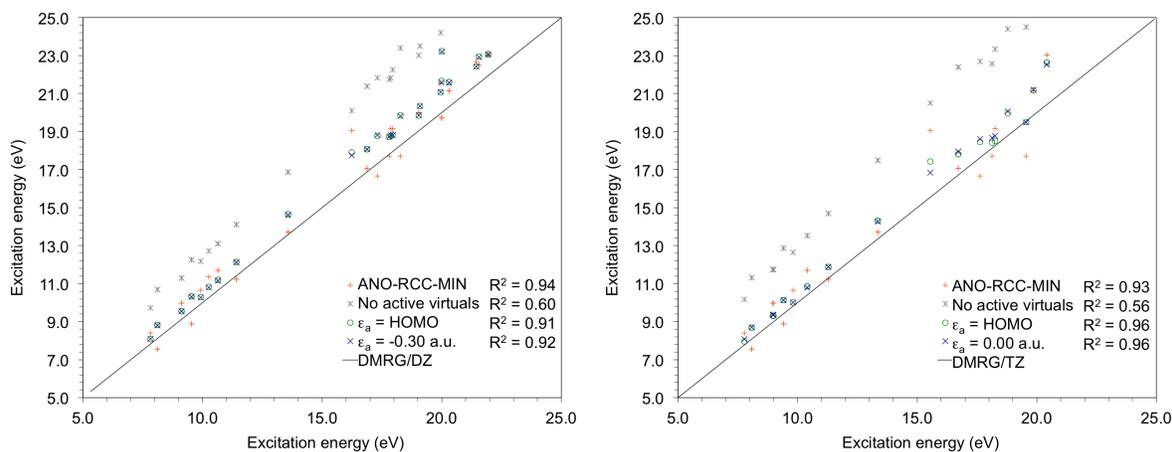

 \includegraphics[scale=0.30]{{{\figures/N2.MIN_DZ.DZ.corrplot}}}
 \includegraphics[scale=0.30]{{{\figures/N2.MIN_TZ.TZ.corrplot}}}
   \caption{Correlation plots for excitation energies of the nitrogen molecule 
with the target ANO-RCC-MIN effective Hamiltonian, compared to the excitation energies in the parent basis (cc-pVDZ (left), cc-pVTZ (right)).  
 (Excitation energies computed using DMRG).
\label{fig:n2.spectrum}}
\end{figure}
\begin{figure}[h!]
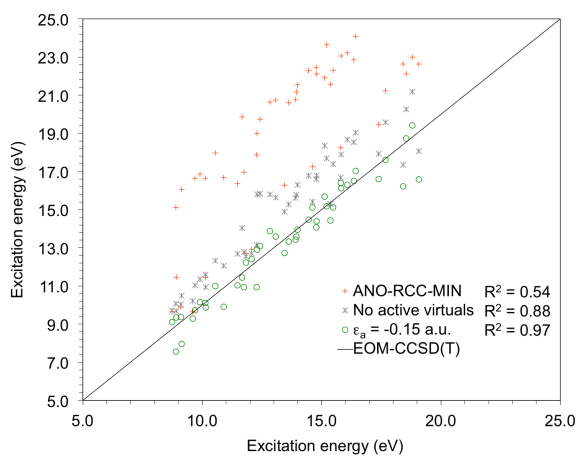

 \includegraphics[scale=0.30]{{{\figures/Ethene.MIN_DZ.DZ.corrplot}}}
   \caption{Correlation plots for excitation energies of the ethylene molecule 
with the target ANO-RCC-MIN effective Hamiltonian, compared to the excitation energies in the parent cc-pVDZ basis. (Excitation energies computed using EOM-CCSD(T)).
\label{fig:ethene.spectrum}}
\end{figure}

Figures~\ref{fig:water.spectrum}, ~\ref{fig:n2.spectrum}, \ref{fig:ethene.spectrum} 
present  correlation plots of the bare and effective Hamiltonian
excitation energies, for  the low-lying excitation energies of the water, nitrogen, and ethylene molecules. (These were generated as the 4 lowest energy states within each
irrep, for each multiplicity). 
We show results for the bare Hamiltonian in DZ and TZ bases, and the effective Hamiltonian 
in a ANO-RCC-MIN basis. 
Statistical data for the excitation energies of water and nitrogen are also given 
in Tables~\ref{tab:water.stats}, \ref{tab:nitrogen.stats}.

All plots clearly show the very substantial improvement brought about by using the 
effective Hamiltonian, rather than the bare Hamiltonian. The
excitation energies (points) for the bare target basis Hamiltonian are quite
far from the exact (parent basis) energies. In the case of water and ethylene they are  too high,
while for nitrogen they are scattered around 
the parent basis results. Folding yields effective
Hamiltonians with excitation energies tightly clustered around the parent results. 

The figures also show the influence of the choice of energies for the active virtuals.
As we saw for the first few excited states above, using no active virtuals
in the effective Hamiltonian construction leads to very poor excitation energies, worse than
using the bare Hamiltonian alone. Using active virtuals and shifting them to
the HOMO energy greatly improves the results, and choosing an optimal 
energy of the active virtuals  improves them  further. This
is also seen through the statistical tables:
for example, for water in the DZ basis, using the effective minimal basis Hamiltonian
reduces the RMSD error from 6.72 eV (bare Hamiltonian) to 1.79 ($\bar{\epsilon}_a$ = HOMO energy) and 1.05 eV ($\bar{\epsilon}_a=-0.3$ eV). 
For nitrogen, as discussed earlier, the minimal basis excitation energies are 
reasonably accurate, and we observe a less dramatic reduction of the RMSD error, 
from 1.24 eV to 1.01 eV ($\bar{\epsilon}_a=-0.3$ eV). 

In explicitly correlated theory, it is common to refer to the effect of including
an explicit F12 correlation factor as increasing the ``zeta'' level of the calculation,
for example, turning a DZ ground-state quality calculation into a QZ quality calculation.
In a similar way, we ask, if we use a very large parent basis, how much does
that increase the ``$\zeta$'' level of the effective Hamiltonian target basis calculation? In Fig.~\ref{fig:water.spectrum}
we further compare the  errors of excitations (measured from
TZ basis excitations) with those from
an effective ANO-RCC-MIN Hamiltonian derived from the TZ parent basis, 
and an explicit (bare Hamiltonian) DZ calculation. We find that 
the canonical transformation approximately achieves one additional
$\zeta$ level of quality.

\begin{table}[ht!]
  \caption{Statistical information for correlation plots of the excitation energies of water.
           $R^2$ and RMSD values.}
  \label{tab:water.stats}\centering
  \begin{threeparttable}
  \begin{tabular}{lcccc}
     \hline
     \multicolumn{1}{c}{Method}  &  $R^2$  & RMSD (eV) &\multicolumn{2}{c}{Average deviations (eV)}\\
     \cline{4-5}
                                 &         &           & Signed & Unsigned\\
     \hline
      Parent=DZ & & & & \\
      \4sp Bare $H$ ANO-RCC-MIN                    & 0.77 &  6.72 & -6.31 & 6.31 \\
      \4sp $\bar{H}$ ANO-RCC-MIN no active    & 0.65 & 10.36 & -9.92 & 9.92 \\
      \4sp $\bar{H}$ ANO-RCC-MIN $\bar{\epsilon}_a=$HOMO  & 0.97 &  1.79 & -1.34 & 1.35 \\
      \4sp $\bar{H}$ ANO-RCC-MIN $\bar{\epsilon}_a=-0.30$ & 0.99 &  1.05 & -0.07 & 0.75 \\[1.0ex]
      Parent=TZ & & & & \\
      \4sp Bare $H$ ANO-RCC-MIN                    & 0.68 &  7.43 & -6.53 & 6.53 \\
      \4sp $\bar{H}$ ANO-RCC-MIN no active    & 0.55 & 12.12 &-11.17 & 11.17\\
      \4sp $\bar{H}$ ANO-RCC-MIN $\bar{\epsilon}_a=$HOMO  & 0.88 &  3.56 & -2.37 & 2.46 \\
      \4sp $\bar{H}$ ANO-RCC-MIN $\bar{\epsilon}_a=-0.30$ & 0.95 &  2.15 & -0.02 & 1.77 \\[1.0ex]
      %
     \hline
  \end{tabular}
  \end{threeparttable}
\end{table}

\begin{table}
  \caption{Statistical information for correlation plots of the excitation energies of nitrogen,
           $R^2$ and RMSD values.}
  \label{tab:nitrogen.stats}\centering
  \begin{threeparttable}
  \begin{tabular}{lcccc}
     \hline
     \multicolumn{1}{c}{Method}  &  $R^2$  & RMSD (eV) &\multicolumn{2}{c}{Average deviations (eV)}\\
     \cline{4-5}
                                 &         &           & Signed & Unsigned\\
     \hline
      Parent=DZ & & & & \\
      \4sp Bare $H$ ANO-RCC-MIN                    & 0.97 &  1.24 & -0.61 & 0.89 \\
      \4sp $\bar{H}$ ANO-RCC-MIN no active    & 0.84 &  4.13 & -3.85 & 3.85 \\
      \4sp $\bar{H}$ ANO-RCC-MIN $\bar{\epsilon}_a=$HOMO  & 0.97 &  1.52 & -1.17 & 1.17 \\
      \4sp $\bar{H}$ ANO-RCC-MIN $\bar{\epsilon}_a=-0.30$ & 0.97 &  1.01 & -0.94 & .0.94\\[1.0ex]
      Parent=TZ & & & & \\
      \4sp Bare $H$ ANO-RCC-MIN                    & 0.94 &  1.31 & -0.54 & 0.99 \\
      \4sp $\bar{H}$ ANO-RCC-MIN no active    & 0.80 &  4.43 & -4.25 & 4.25 \\
      \4sp $\bar{H}$ ANO-RCC-MIN $\bar{\epsilon}_a=$HOMO  & 0.98 &  0.99 & -0.80 & 0.80 \\
      \4sp $\bar{H}$ ANO-RCC-MIN $\bar{\epsilon}_a=-0.30$ & 0.98 &  0.95 & -0.75 & 0.77 \\[1.0ex]
      %
     \hline
  \end{tabular}
  \end{threeparttable}
\end{table}

\subsection{Potential energy curves}

\begin{figure}
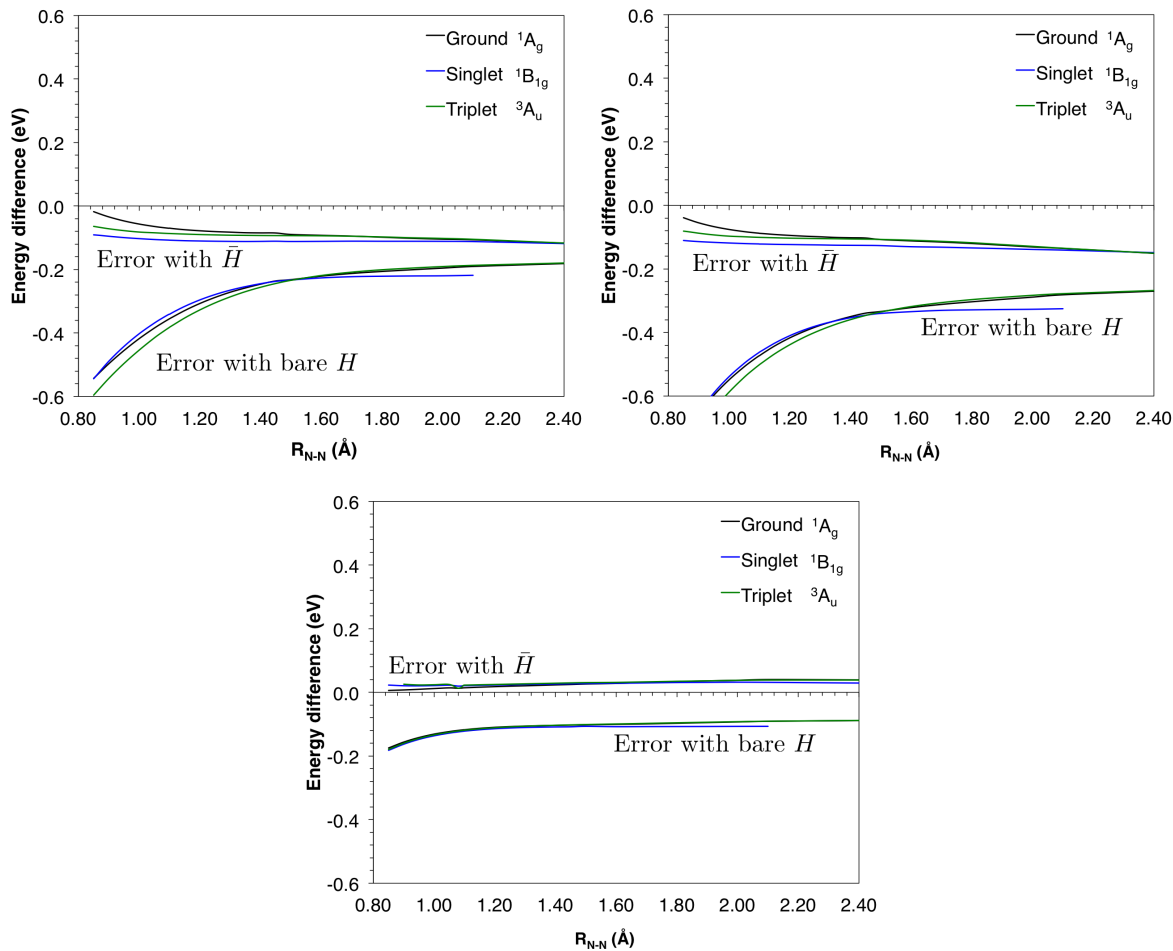

 \includegraphics[scale=0.35]{{{\figures/N2.curves.MIN_DZ.DZ}}} 
 \includegraphics[scale=0.35]{{{\figures/N2.curves.MIN_TZ.TZ}}} \\
 \includegraphics[scale=0.35]{{{\figures/N2.curves.DZ_TZ.TZ}}}
   \caption{Parallelity errors in the ground, first singlet excited, and first triplet excited state using the bare Hamiltonian $H$ and the effective Hamiltonian $\bar{H}$. The curves are generated
using CASSCF(6,6)+CTSD. Top: parent basis=cc-pVDZ, left: target basis=ANO-RCC-MIN, right: target basis=ANO-RCC-MIN, Bottom: parent basis=cc-pVTZ, target basis=cc-pVDZ. The reference basis for each figure is the parent basis.}
\label{fig:N2.curves}
\end{figure}

As a further test of our effective Hamiltonians, we now study
how they perform in describing potential energy curves in 
a minimal (or small) basis. We first consider the
nitrogen dimer. In Figure~\ref{fig:N2.curves} we show the errors in the ground, first singlet excited, and first triplet excited state potential energy curves computed
by diagonalizing the effective Hamiltonian
in the ANO-RCC-MIN basis, as compared to the parent basis. (The 
 effective Hamiltonian is derived for the singlet ground-state). For comparison we also show
 errors of the curves computed using the bare Hamiltonian in the ANO-RCC-MIN basis. 
For each basis, the curves were computed using {full valence complete active space self-consistent field
followed by a canonical transformation with singles and doubles excitations dynamic
correlation treatment (CASSCF+CTSD).}


As expected, the bare Hamiltonian potential curves in the minimal basis
display very large errors in the compressed bond region, due
to the lack of a second shell to describe polarizing effects. In contrast,
the effective Hamiltonian yields much smaller errors across the curve (roughly
half the error at long bond-lengths, and much higher accuracy at short bond lengths). But more
importantly, the errors are very parallel across the entire range
of bond-lengths, and consistent between all the states.

As a more challenging probe of the quality of the chemistry 
that can be described by two-particle Hamiltonians in a minimal basis,
we now consider the chromium dimer binding curve.  
The chromium dimer has been the subject of numerous quantum chemical
studies due to the difficulties in obtaining a potential curve
of even qualitatively correct shape~\cite{roos1995multiconfigurational,dachsel1999multireference,celani2004cipt2,kurashige2011second}. It is well known that, in addition
to the strong correlation arising from the spin coupling
of the chromium  d electrons, very large basis sets
are needed~\cite{roos2003ground,kurashige2011second}. 
Thus constructing a two-particle effective Hamiltonian
for a minimal basis description is a serious challenge.


As a parent basis, we used high quality ANO-RCC bases of \dz, \tz, \qz\ quality~\cite{roos2005new}
supplemented with an additional set of $d$-functions taken from
the next $\zeta$ in the series. This yields the following basis set labels and structure: ANO-RCC-DZP+$d$ (21s15p10d6f/5s3p3d1f), ANO-RCC-TZP+$d$ (21s15p10d6f4g/6s4p4d2f1g), and ANO-RCC-QZP+$d$ (21s15p10d6f4g2h/7s5p5d3f2g1h). We used the valence (12,12) active space in the CASSCF calculation.  We first tried folding to a target ANO-RCC-MIN (21s15p10d/4s2p1d) basis. While
this yielded a bound potential, the binding energy was $111.4$~{kcal mol$^{-1}$} ($4.83$~eV), several times
too large, showing that, at least using our procedure, a reasonable effective two-particle  Hamiltonian in a strict minimal basis
cannot be constructed. We next tried folding to a slightly larger ANO-RCC-MIN+$d$ basis, 
where an additional
d shell (taken from the {\dz} basis (21s15p10d/4s2p2d)) was included in the target basis 
to capture polarization effects. The CASSCF density
matrix was used in the $(1,2)$ construction of $\bar{H}$ as the Hartree-Fock potential
energy surface has an inconvenient curve crossing. 

\begin{figure}[h!]
 \includegraphics[scale=0.5]{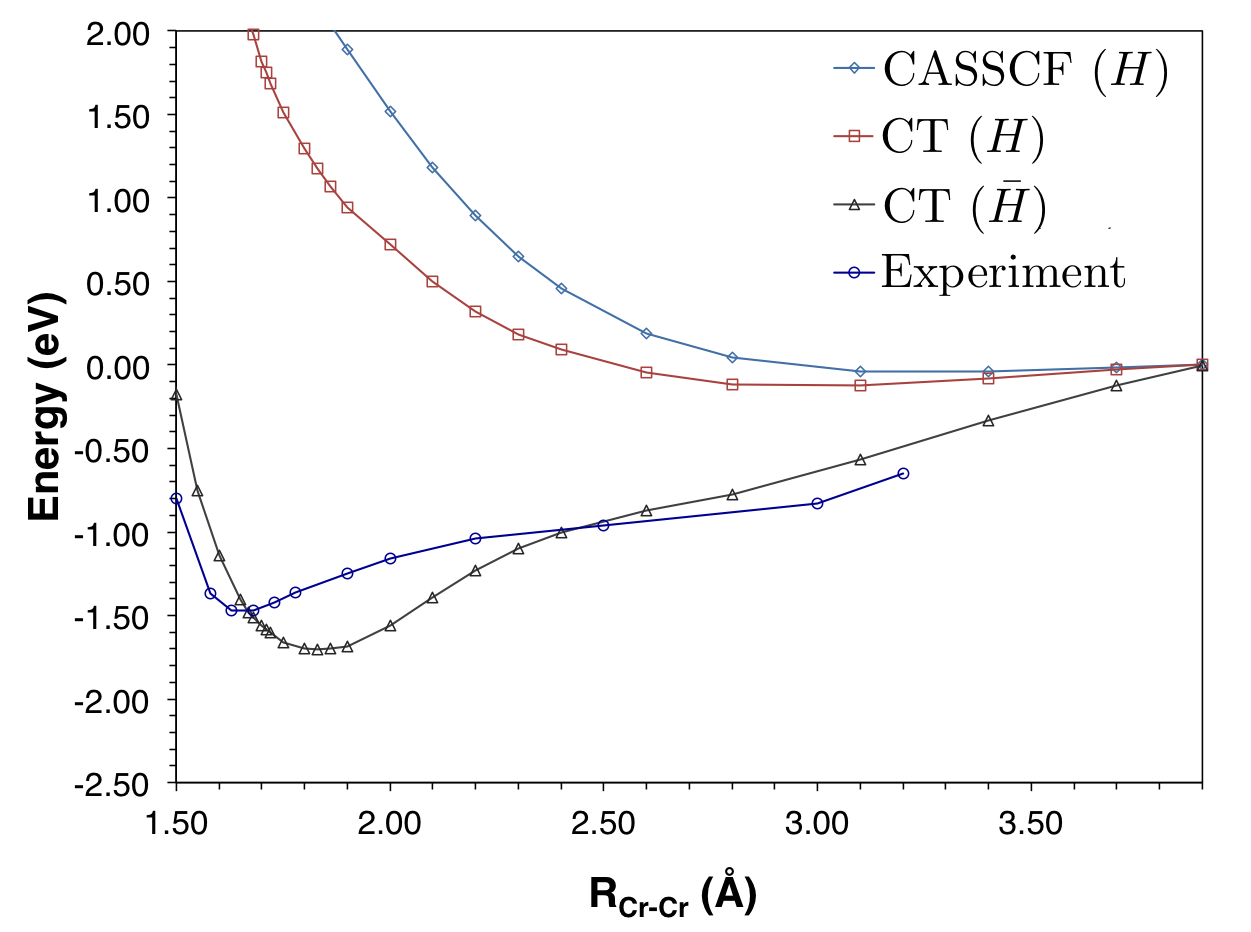}
   \caption{Potential energy curves for the chromium dimer for the different methods shown using the
            ANO-RCC-MIN+$d$ basis set.
External basis set is \anoxdz{Q}.}
\label{fig:cr2.curves}
\end{figure}

The experimental curve is shown in Fig~\ref{fig:cr2.curves}. The bare Hamiltonian ANO-RCC-MIN+$d$ basis results are shown using CASSCF(12,12) and
CASSCF(12,12) plus CTSD. Neither of these curves show any binding, as expected in a minimal basis. 
We see, however, that the CTSD calculations with the effective Hamiltonian in the 
ANO-RCC-MIN+$d$ basis yield a nicely bound potential energy curve
of a very similar shape and depth to experiment! 

The  values for the binding energy, equilibrium distance, and spectroscopic constants, obtained using  the aforementioned constructed ANO-RCC parent bases, DZP+$d$, TZP+$d$, QZP+$d$, and the target ANO-RCC-MIN+$d$ basis, are given in 
Table~\ref{tab:cr2.req_de}.
Even using the \anoxdz{D}\ external basis we recover binding in the effective ANO-RCC-MIN+$d$ Hamiltonian.
Folding down from the largest QZ parent basis, we obtain for our effective ANO-RCC-MIN+$d$ Hamiltonian 
 an R$_{eq}$ of $1.83$~{\AA} and $D_e$ of
$1.70$~eV. This compares quite favorably with experiment. Overall, this shows that it is 
possible to construct an effective Hamiltonian
to describe even this very difficult case of the binding of the chromium dimer, so long as 
the minimal basis is slightly expanded. We understand this because the minimal basis for Cr$_2$ with a (12,12) active space does not leave any virtual orbitals to be used in a
correlated calculation after the folding procedure. In this
strongly correlated system, the role of the extra set of 
d-functions is to give 10 virtual
orbitals outside the active space that help to relax the orbitals.

\begin{table}[ht!]
  \caption[R$_{eq}$ and D$_{e}$ for Cr$_2$ with $\cth_{corr.}$]
          { Equilibrium bond distances and dissociation energies for
            the chromium dimer using an effective Hamiltonian in a minimal + $d$ basis,
            folded down from increasingly large basis sets (DZP, TZP, QZP), and
            for different values of the active virtual energy $\bar{\epsilon}_a$.}
  \label{tab:cr2.req_de}\centering
  \begin{threeparttable}
  \begin{tabular}{ccc}
     \hline
      Basis  &  R$_{eq}$ ({\AA}) & $D_e$ (eV) \\
     \hline
     MIN+$d$ (DZP+$d$)  & &  \\
     $\bar{\epsilon}_a$=0.00 & 2.20 & 1.55 \\
     $\bar{\epsilon}_a$=0.05 & 2.10 & 1.66 \\
     $\bar{\epsilon}_a$=0.10 & 1.90 & 1.84 \\[1.0ex]
     MIN+$d$ (TZP+$d$)  & &  \\
     $\bar{\epsilon}_a$=0.00 & 2.40 & 0.75 \\
     $\bar{\epsilon}_a$=0.04 & 2.40 & 0.78 \\
     $\bar{\epsilon}_a$=0.05 & 2.40 & 0.79 \\[1.0ex]
     MIN+$d$ (QZP+$d$)  & &  \\
     $\bar{\epsilon}_a$=0.00 & 1.83 & 1.70 \\
     Experiment    & 1.679$^a$&  1.47$^b$\\
                   &      & 1.45$^c$ \\
                   &      & 1.56$^d$ \\
     \hline
  \end{tabular}
  \begin{tablenotes}
     \item[a] Ref.~\cite{bondybey1983electronic}
     \item[b] Ref.~\cite{casey1993negative}
     \item[c] Ref.~\cite{hilpert1989ber}
     \item[d] Ref.~\cite{su1993bond}
  \end{tablenotes}
  \end{threeparttable}
\end{table}

\section{Conclusions}
\label{sec:conclusions}

In this work we asked whether a simple
canonical transformation, using a single-reference-like
modified second-order perturbation theory formula, yields an effective
Hamiltonian in a minimal (or very small) basis with qualitatively 
correct chemistry.
As we saw, the answer is in the affirmative: such minimal basis set 
effective Hamiltonians 
give qualitatively correct excitation energies and binding curves 
in water, nitrogen, ethylene, and even the chromium dimer!

Effective Hamiltonians 
formally provide a conceptual link between quantitative and qualitative
reasoning. The simple nature of the construction here
means that we can now  derive accurate effective Hamiltonians
 for complex systems in practice, including systems with transition metals,  
and for correlated electrons in the condensed phase, where models are essential
not only for interpretation but for practical computation.
Intriguingly,  our  various calculations suggests that these simple effective Hamiltonians
may even be {\it quantitatively} accurate. The technique here thus further provides the possibility of
 very low cost (that is lower than multireference perturbation theory) 
treatment of dynamic correlation in  challenging multireference 
quantum chemistry.

\bibliography{FoldToMinBasis}

\providecommand{\latin}[1]{#1}
\providecommand*\mcitethebibliography{\thebibliography}
\csname @ifundefined\endcsname{endmcitethebibliography}
  {\let\endmcitethebibliography\endthebibliography}{}
\begin{mcitethebibliography}{55}
\providecommand*\natexlab[1]{#1}
\providecommand*\mciteSetBstSublistMode[1]{}
\providecommand*\mciteSetBstMaxWidthForm[2]{}
\providecommand*\mciteBstWouldAddEndPuncttrue
  {\def\EndOfBibitem{\unskip.}}
\providecommand*\mciteBstWouldAddEndPunctfalse
  {\let\EndOfBibitem\relax}
\providecommand*\mciteSetBstMidEndSepPunct[3]{}
\providecommand*\mciteSetBstSublistLabelBeginEnd[3]{}
\providecommand*\EndOfBibitem{}
\mciteSetBstSublistMode{f}
\mciteSetBstMaxWidthForm{subitem}{(\alph{mcitesubitemcount})}
\mciteSetBstSublistLabelBeginEnd
  {\mcitemaxwidthsubitemform\space}
  {\relax}
  {\relax}

\bibitem[Polyansky \latin{et~al.}(2003)Polyansky, Cs{\'a}sz{\'a}r, Shirin,
  Zobov, Barletta, Tennyson, Schwenke, and Knowles]{polyansky2003high}
Polyansky,~O.~L.; Cs{\'a}sz{\'a}r,~A.~G.; Shirin,~S.~V.; Zobov,~N.~F.;
  Barletta,~P.; Tennyson,~J.; Schwenke,~D.~W.; Knowles,~P.~J. \emph{Science}
  \textbf{2003}, \emph{299}, 539--542\relax
\mciteBstWouldAddEndPuncttrue
\mciteSetBstMidEndSepPunct{\mcitedefaultmidpunct}
{\mcitedefaultendpunct}{\mcitedefaultseppunct}\relax
\EndOfBibitem
\bibitem[Karton \latin{et~al.}(2006)Karton, Rabinovich, Martin, and
  Ruscic]{karton2006w4}
Karton,~A.; Rabinovich,~E.; Martin,~J.~M.; Ruscic,~B. \emph{The Journal of
  chemical physics} \textbf{2006}, \emph{125}, 144108\relax
\mciteBstWouldAddEndPuncttrue
\mciteSetBstMidEndSepPunct{\mcitedefaultmidpunct}
{\mcitedefaultendpunct}{\mcitedefaultseppunct}\relax
\EndOfBibitem
\bibitem[Tajti \latin{et~al.}(2004)Tajti, Szalay, Cs{\'a}sz{\'a}r, K{\'a}llay,
  Gauss, Valeev, Flowers, V{\'a}zquez, and Stanton]{tajti2004heat}
Tajti,~A.; Szalay,~P.~G.; Cs{\'a}sz{\'a}r,~A.~G.; K{\'a}llay,~M.; Gauss,~J.;
  Valeev,~E.~F.; Flowers,~B.~A.; V{\'a}zquez,~J.; Stanton,~J.~F. \emph{The
  Journal of chemical physics} \textbf{2004}, \emph{121}, 11599--11613\relax
\mciteBstWouldAddEndPuncttrue
\mciteSetBstMidEndSepPunct{\mcitedefaultmidpunct}
{\mcitedefaultendpunct}{\mcitedefaultseppunct}\relax
\EndOfBibitem
\bibitem[Sharma \latin{et~al.}(2014)Sharma, Yanai, Booth, Umrigar, and
  Chan]{sharma2014spectroscopic}
Sharma,~S.; Yanai,~T.; Booth,~G.~H.; Umrigar,~C.; Chan,~G. K.-L. \emph{The
  Journal of chemical physics} \textbf{2014}, \emph{140}, 104112\relax
\mciteBstWouldAddEndPuncttrue
\mciteSetBstMidEndSepPunct{\mcitedefaultmidpunct}
{\mcitedefaultendpunct}{\mcitedefaultseppunct}\relax
\EndOfBibitem
\bibitem[Yang \latin{et~al.}(2014)Yang, Hu, Usvyat, Matthews, Sch{\"u}tz, and
  Chan]{yang2014ab}
Yang,~J.; Hu,~W.; Usvyat,~D.; Matthews,~D.; Sch{\"u}tz,~M.; Chan,~G. K.-L.
  \emph{Science} \textbf{2014}, \emph{345}, 640--643\relax
\mciteBstWouldAddEndPuncttrue
\mciteSetBstMidEndSepPunct{\mcitedefaultmidpunct}
{\mcitedefaultendpunct}{\mcitedefaultseppunct}\relax
\EndOfBibitem
\bibitem[Blaizot and Ripka(1986)Blaizot, and Ripka]{blaizot1986quantum}
Blaizot,~J.-P.; Ripka,~G. \emph{Quantum theory of finite systems}; Mit Press
  Cambridge, 1986; Vol.~3\relax
\mciteBstWouldAddEndPuncttrue
\mciteSetBstMidEndSepPunct{\mcitedefaultmidpunct}
{\mcitedefaultendpunct}{\mcitedefaultseppunct}\relax
\EndOfBibitem
\bibitem[Onida \latin{et~al.}(2002)Onida, Reining, and
  Rubio]{onida2002electronic}
Onida,~G.; Reining,~L.; Rubio,~A. \emph{Reviews of Modern Physics}
  \textbf{2002}, \emph{74}, 601\relax
\mciteBstWouldAddEndPuncttrue
\mciteSetBstMidEndSepPunct{\mcitedefaultmidpunct}
{\mcitedefaultendpunct}{\mcitedefaultseppunct}\relax
\EndOfBibitem
\bibitem[Werner and Millis(2010)Werner, and Millis]{werner2010dynamical}
Werner,~P.; Millis,~A.~J. \emph{Physical review letters} \textbf{2010},
  \emph{104}, 146401\relax
\mciteBstWouldAddEndPuncttrue
\mciteSetBstMidEndSepPunct{\mcitedefaultmidpunct}
{\mcitedefaultendpunct}{\mcitedefaultseppunct}\relax
\EndOfBibitem
\bibitem[Kemble(1937)]{kemble1937fundamental}
Kemble,~E.~C. \emph{The fundamental principles of quantum mechanics}; Dover,
  1937\relax
\mciteBstWouldAddEndPuncttrue
\mciteSetBstMidEndSepPunct{\mcitedefaultmidpunct}
{\mcitedefaultendpunct}{\mcitedefaultseppunct}\relax
\EndOfBibitem
\bibitem[Brandow(1967)]{brandow1967linked}
Brandow,~B.~H. \emph{Reviews of Modern Physics} \textbf{1967}, \emph{39},
  771\relax
\mciteBstWouldAddEndPuncttrue
\mciteSetBstMidEndSepPunct{\mcitedefaultmidpunct}
{\mcitedefaultendpunct}{\mcitedefaultseppunct}\relax
\EndOfBibitem
\bibitem[Westhaus(1973)]{westhaus1973cluster}
Westhaus,~P. \emph{International Journal of Quantum Chemistry} \textbf{1973},
  \emph{7}, 463--477\relax
\mciteBstWouldAddEndPuncttrue
\mciteSetBstMidEndSepPunct{\mcitedefaultmidpunct}
{\mcitedefaultendpunct}{\mcitedefaultseppunct}\relax
\EndOfBibitem
\bibitem[Freed(1974)]{freed1974theoretical}
Freed,~K.~F. \emph{The Journal of Chemical Physics} \textbf{1974}, \emph{60},
  1765--1788\relax
\mciteBstWouldAddEndPuncttrue
\mciteSetBstMidEndSepPunct{\mcitedefaultmidpunct}
{\mcitedefaultendpunct}{\mcitedefaultseppunct}\relax
\EndOfBibitem
\bibitem[Iwata and Freed(1976)Iwata, and Freed]{iwata1976nonclassical}
Iwata,~S.; Freed,~K.~F. \emph{The Journal of Chemical Physics} \textbf{1976},
  \emph{65}, 1071--1088\relax
\mciteBstWouldAddEndPuncttrue
\mciteSetBstMidEndSepPunct{\mcitedefaultmidpunct}
{\mcitedefaultendpunct}{\mcitedefaultseppunct}\relax
\EndOfBibitem
\bibitem[Durand and Malrieu(2009)Durand, and Malrieu]{durand2009effective}
Durand,~P.; Malrieu,~J.-P. \emph{Advances in chemical Physics} \textbf{2009},
  321--412\relax
\mciteBstWouldAddEndPuncttrue
\mciteSetBstMidEndSepPunct{\mcitedefaultmidpunct}
{\mcitedefaultendpunct}{\mcitedefaultseppunct}\relax
\EndOfBibitem
\bibitem[Wegner(1994)]{wegner1994flow}
Wegner,~F. \emph{Annalen der physik} \textbf{1994}, \emph{506}, 77--91\relax
\mciteBstWouldAddEndPuncttrue
\mciteSetBstMidEndSepPunct{\mcitedefaultmidpunct}
{\mcitedefaultendpunct}{\mcitedefaultseppunct}\relax
\EndOfBibitem
\bibitem[G{\l}azek and Wilson(1993)G{\l}azek, and
  Wilson]{glazek1993renormalization}
G{\l}azek,~S.~D.; Wilson,~K.~G. \emph{Physical Review D} \textbf{1993},
  \emph{48}, 5863\relax
\mciteBstWouldAddEndPuncttrue
\mciteSetBstMidEndSepPunct{\mcitedefaultmidpunct}
{\mcitedefaultendpunct}{\mcitedefaultseppunct}\relax
\EndOfBibitem
\bibitem[White(2002)]{white2002numerical}
White,~S.~R. \emph{The Journal of chemical physics} \textbf{2002}, \emph{117},
  7472--7482\relax
\mciteBstWouldAddEndPuncttrue
\mciteSetBstMidEndSepPunct{\mcitedefaultmidpunct}
{\mcitedefaultendpunct}{\mcitedefaultseppunct}\relax
\EndOfBibitem
\bibitem[Bartlett and Musia{\l}(2007)Bartlett, and
  Musia{\l}]{bartlett2007coupled}
Bartlett,~R.~J.; Musia{\l},~M. \emph{Reviews of Modern Physics} \textbf{2007},
  \emph{79}, 291\relax
\mciteBstWouldAddEndPuncttrue
\mciteSetBstMidEndSepPunct{\mcitedefaultmidpunct}
{\mcitedefaultendpunct}{\mcitedefaultseppunct}\relax
\EndOfBibitem
\bibitem[Stanton and Bartlett(1993)Stanton, and Bartlett]{stanton1993equation}
Stanton,~J.~F.; Bartlett,~R.~J. \emph{The Journal of chemical physics}
  \textbf{1993}, \emph{98}, 7029--7039\relax
\mciteBstWouldAddEndPuncttrue
\mciteSetBstMidEndSepPunct{\mcitedefaultmidpunct}
{\mcitedefaultendpunct}{\mcitedefaultseppunct}\relax
\EndOfBibitem
\bibitem[Lyakh \latin{et~al.}(2011)Lyakh, Musia{\l}, Lotrich, and
  Bartlett]{lyakh2011multireference}
Lyakh,~D.~I.; Musia{\l},~M.; Lotrich,~V.~F.; Bartlett,~R.~J. \emph{Chemical
  reviews} \textbf{2011}, \emph{112}, 182--243\relax
\mciteBstWouldAddEndPuncttrue
\mciteSetBstMidEndSepPunct{\mcitedefaultmidpunct}
{\mcitedefaultendpunct}{\mcitedefaultseppunct}\relax
\EndOfBibitem
\bibitem[Yanai and Chan(2006)Yanai, and Chan]{yanai2006canonical}
Yanai,~T.; Chan,~G. K.-L. \emph{The Journal of chemical physics} \textbf{2006},
  \emph{124}, 194106\relax
\mciteBstWouldAddEndPuncttrue
\mciteSetBstMidEndSepPunct{\mcitedefaultmidpunct}
{\mcitedefaultendpunct}{\mcitedefaultseppunct}\relax
\EndOfBibitem
\bibitem[Yanai and Chan(2007)Yanai, and Chan]{yanai2007canonical}
Yanai,~T.; Chan,~G. K.-L. \emph{The Journal of chemical physics} \textbf{2007},
  \emph{127}, 104107\relax
\mciteBstWouldAddEndPuncttrue
\mciteSetBstMidEndSepPunct{\mcitedefaultmidpunct}
{\mcitedefaultendpunct}{\mcitedefaultseppunct}\relax
\EndOfBibitem
\bibitem[Neuscamman \latin{et~al.}(2010)Neuscamman, Yanai, and
  Chan]{neuscamman2010strongly}
Neuscamman,~E.; Yanai,~T.; Chan,~G. K.-L. \emph{The Journal of chemical
  physics} \textbf{2010}, \emph{132}, 024106\relax
\mciteBstWouldAddEndPuncttrue
\mciteSetBstMidEndSepPunct{\mcitedefaultmidpunct}
{\mcitedefaultendpunct}{\mcitedefaultseppunct}\relax
\EndOfBibitem
\bibitem[Neuscamman \latin{et~al.}(2010)Neuscamman, Yanai, and
  Chan]{neuscamman2010review}
Neuscamman,~E.; Yanai,~T.; Chan,~G. K.-L. \emph{International Reviews in
  Physical Chemistry} \textbf{2010}, \emph{29}, 231--271\relax
\mciteBstWouldAddEndPuncttrue
\mciteSetBstMidEndSepPunct{\mcitedefaultmidpunct}
{\mcitedefaultendpunct}{\mcitedefaultseppunct}\relax
\EndOfBibitem
\bibitem[Mazziotti(2006)]{mazziotti2006anti}
Mazziotti,~D.~A. \emph{Physical review letters} \textbf{2006}, \emph{97},
  143002\relax
\mciteBstWouldAddEndPuncttrue
\mciteSetBstMidEndSepPunct{\mcitedefaultmidpunct}
{\mcitedefaultendpunct}{\mcitedefaultseppunct}\relax
\EndOfBibitem
\bibitem[Mazziotti(2007)]{mazziotti2007anti}
Mazziotti,~D.~A. \emph{Physical Review A} \textbf{2007}, \emph{75},
  022505\relax
\mciteBstWouldAddEndPuncttrue
\mciteSetBstMidEndSepPunct{\mcitedefaultmidpunct}
{\mcitedefaultendpunct}{\mcitedefaultseppunct}\relax
\EndOfBibitem
\bibitem[Evangelista(2014)]{evangelista2014driven}
Evangelista,~F.~A. \emph{arXiv preprint arXiv:1406.0114} \textbf{2014}, \relax
\mciteBstWouldAddEndPunctfalse
\mciteSetBstMidEndSepPunct{\mcitedefaultmidpunct}
{}{\mcitedefaultseppunct}\relax
\EndOfBibitem
\bibitem[Yanai and Shiozaki(2012)Yanai, and Shiozaki]{yanai2012canonical}
Yanai,~T.; Shiozaki,~T. \emph{The Journal of chemical physics} \textbf{2012},
  \emph{136}, 084107\relax
\mciteBstWouldAddEndPuncttrue
\mciteSetBstMidEndSepPunct{\mcitedefaultmidpunct}
{\mcitedefaultendpunct}{\mcitedefaultseppunct}\relax
\EndOfBibitem
\bibitem[Valeev(2004)]{valeev2004improving}
Valeev,~E.~F. \emph{Chemical physics letters} \textbf{2004}, \emph{395},
  190--195\relax
\mciteBstWouldAddEndPuncttrue
\mciteSetBstMidEndSepPunct{\mcitedefaultmidpunct}
{\mcitedefaultendpunct}{\mcitedefaultseppunct}\relax
\EndOfBibitem
\bibitem[Roos \latin{et~al.}(2005)Roos, Lindh, Malmqvist, Veryazov, and
  Widmark]{roos2005new}
Roos,~B.~O.; Lindh,~R.; Malmqvist,~P.-{\AA}.; Veryazov,~V.; Widmark,~P.-O.
  \emph{The Journal of Physical Chemistry A} \textbf{2005}, \emph{109},
  6575--6579\relax
\mciteBstWouldAddEndPuncttrue
\mciteSetBstMidEndSepPunct{\mcitedefaultmidpunct}
{\mcitedefaultendpunct}{\mcitedefaultseppunct}\relax
\EndOfBibitem
\bibitem[Dunning(1989)]{dunning1989basis}
Dunning,~T.~H. \emph{The Journal of chemical physics} \textbf{1989}, \emph{90},
  1007\relax
\mciteBstWouldAddEndPuncttrue
\mciteSetBstMidEndSepPunct{\mcitedefaultmidpunct}
{\mcitedefaultendpunct}{\mcitedefaultseppunct}\relax
\EndOfBibitem
\bibitem[Kutzelnigg(1979)]{kutzelnigg1979generalized}
Kutzelnigg,~W. \emph{Chemical Physics Letters} \textbf{1979}, \emph{64},
  383--387\relax
\mciteBstWouldAddEndPuncttrue
\mciteSetBstMidEndSepPunct{\mcitedefaultmidpunct}
{\mcitedefaultendpunct}{\mcitedefaultseppunct}\relax
\EndOfBibitem
\bibitem[Neuscamman \latin{et~al.}(2009)Neuscamman, Yanai, and
  Chan]{neuscamman2009quadratic}
Neuscamman,~E.; Yanai,~T.; Chan,~G. K.-L. \emph{The Journal of chemical
  physics} \textbf{2009}, \emph{130}, 124102\relax
\mciteBstWouldAddEndPuncttrue
\mciteSetBstMidEndSepPunct{\mcitedefaultmidpunct}
{\mcitedefaultendpunct}{\mcitedefaultseppunct}\relax
\EndOfBibitem
\bibitem[Kutzelnigg and Mukherjee(1997)Kutzelnigg, and
  Mukherjee]{kutzelnigg1997normal}
Kutzelnigg,~W.; Mukherjee,~D. \emph{The Journal of chemical physics}
  \textbf{1997}, \emph{107}, 432--449\relax
\mciteBstWouldAddEndPuncttrue
\mciteSetBstMidEndSepPunct{\mcitedefaultmidpunct}
{\mcitedefaultendpunct}{\mcitedefaultseppunct}\relax
\EndOfBibitem
\bibitem[Mazziotti(1998)]{mazziotti1998approximate}
Mazziotti,~D.~A. \emph{Chemical physics letters} \textbf{1998}, \emph{289},
  419--427\relax
\mciteBstWouldAddEndPuncttrue
\mciteSetBstMidEndSepPunct{\mcitedefaultmidpunct}
{\mcitedefaultendpunct}{\mcitedefaultseppunct}\relax
\EndOfBibitem
\bibitem[Mazziotti(1998)]{mazziotti1998contracted}
Mazziotti,~D.~A. \emph{Physical Review A} \textbf{1998}, \emph{57}, 4219\relax
\mciteBstWouldAddEndPuncttrue
\mciteSetBstMidEndSepPunct{\mcitedefaultmidpunct}
{\mcitedefaultendpunct}{\mcitedefaultseppunct}\relax
\EndOfBibitem
\bibitem[Kutzelnigg and Mukherjee(1999)Kutzelnigg, and
  Mukherjee]{kutzelnigg1999cumulant}
Kutzelnigg,~W.; Mukherjee,~D. \emph{Journal of Chemical Physics} \textbf{1999},
  \emph{110}, 2800--2809\relax
\mciteBstWouldAddEndPuncttrue
\mciteSetBstMidEndSepPunct{\mcitedefaultmidpunct}
{\mcitedefaultendpunct}{\mcitedefaultseppunct}\relax
\EndOfBibitem
\bibitem[Shamasundar(2009)]{shamasundar2009cumulant}
Shamasundar,~K. \emph{The Journal of chemical physics} \textbf{2009},
  \emph{131}, 174109\relax
\mciteBstWouldAddEndPuncttrue
\mciteSetBstMidEndSepPunct{\mcitedefaultmidpunct}
{\mcitedefaultendpunct}{\mcitedefaultseppunct}\relax
\EndOfBibitem
\bibitem[Banerjee and Simons(1981)Banerjee, and Simons]{banerjee1981coupled}
Banerjee,~A.; Simons,~J. \emph{International Journal of Quantum Chemistry}
  \textbf{1981}, \emph{19}, 207--216\relax
\mciteBstWouldAddEndPuncttrue
\mciteSetBstMidEndSepPunct{\mcitedefaultmidpunct}
{\mcitedefaultendpunct}{\mcitedefaultseppunct}\relax
\EndOfBibitem
\bibitem[Evangelista and Gauss(2011)Evangelista, and
  Gauss]{evangelista2011orbital}
Evangelista,~F.~A.; Gauss,~J. \emph{The Journal of chemical physics}
  \textbf{2011}, \emph{134}, 114102\relax
\mciteBstWouldAddEndPuncttrue
\mciteSetBstMidEndSepPunct{\mcitedefaultmidpunct}
{\mcitedefaultendpunct}{\mcitedefaultseppunct}\relax
\EndOfBibitem
\bibitem[Hanauer and K{\"o}hn(2011)Hanauer, and K{\"o}hn]{hanauer2011pilot}
Hanauer,~M.; K{\"o}hn,~A. \emph{The Journal of chemical physics} \textbf{2011},
  \emph{134}, 204111\relax
\mciteBstWouldAddEndPuncttrue
\mciteSetBstMidEndSepPunct{\mcitedefaultmidpunct}
{\mcitedefaultendpunct}{\mcitedefaultseppunct}\relax
\EndOfBibitem
\bibitem[Datta \latin{et~al.}(2011)Datta, Kong, and Nooijen]{datta2011state}
Datta,~D.; Kong,~L.; Nooijen,~M. \emph{The Journal of chemical physics}
  \textbf{2011}, \emph{134}, 214116\relax
\mciteBstWouldAddEndPuncttrue
\mciteSetBstMidEndSepPunct{\mcitedefaultmidpunct}
{\mcitedefaultendpunct}{\mcitedefaultseppunct}\relax
\EndOfBibitem
\bibitem[Ten-No(2004)]{ten2004explicitly}
Ten-No,~S. \emph{The Journal of chemical physics} \textbf{2004}, \emph{121},
  117--129\relax
\mciteBstWouldAddEndPuncttrue
\mciteSetBstMidEndSepPunct{\mcitedefaultmidpunct}
{\mcitedefaultendpunct}{\mcitedefaultseppunct}\relax
\EndOfBibitem
\bibitem[Ked{\v{z}}uch \latin{et~al.}(2005)Ked{\v{z}}uch, Milko, and
  Noga]{kedvzuch2005alternative}
Ked{\v{z}}uch,~S.; Milko,~M.; Noga,~J. \emph{International journal of quantum
  chemistry} \textbf{2005}, \emph{105}, 929--936\relax
\mciteBstWouldAddEndPuncttrue
\mciteSetBstMidEndSepPunct{\mcitedefaultmidpunct}
{\mcitedefaultendpunct}{\mcitedefaultseppunct}\relax
\EndOfBibitem
\bibitem[Kendall \latin{et~al.}(1992)Kendall, Dunning~Jr, and
  Harrison]{kendall1992electron}
Kendall,~R.~A.; Dunning~Jr,~T.~H.; Harrison,~R.~J. \emph{The Journal of
  chemical physics} \textbf{1992}, \emph{96}, 6796--6806\relax
\mciteBstWouldAddEndPuncttrue
\mciteSetBstMidEndSepPunct{\mcitedefaultmidpunct}
{\mcitedefaultendpunct}{\mcitedefaultseppunct}\relax
\EndOfBibitem
\bibitem[Roos and Andersson(1995)Roos, and
  Andersson]{roos1995multiconfigurational}
Roos,~B.~O.; Andersson,~K. \emph{Chemical physics letters} \textbf{1995},
  \emph{245}, 215--223\relax
\mciteBstWouldAddEndPuncttrue
\mciteSetBstMidEndSepPunct{\mcitedefaultmidpunct}
{\mcitedefaultendpunct}{\mcitedefaultseppunct}\relax
\EndOfBibitem
\bibitem[Dachsel \latin{et~al.}(1999)Dachsel, Harrison, and
  Dixon]{dachsel1999multireference}
Dachsel,~H.; Harrison,~R.~J.; Dixon,~D.~A. \emph{The Journal of Physical
  Chemistry A} \textbf{1999}, \emph{103}, 152--155\relax
\mciteBstWouldAddEndPuncttrue
\mciteSetBstMidEndSepPunct{\mcitedefaultmidpunct}
{\mcitedefaultendpunct}{\mcitedefaultseppunct}\relax
\EndOfBibitem
\bibitem[Celani \latin{et~al.}(2004)Celani, Stoll, Werner, and
  Knowles]{celani2004cipt2}
Celani,~P.; Stoll,~H.; Werner,~H.-J.; Knowles,~P. \emph{Molecular Physics}
  \textbf{2004}, \emph{102}, 2369--2379\relax
\mciteBstWouldAddEndPuncttrue
\mciteSetBstMidEndSepPunct{\mcitedefaultmidpunct}
{\mcitedefaultendpunct}{\mcitedefaultseppunct}\relax
\EndOfBibitem
\bibitem[Kurashige and Yanai(2011)Kurashige, and Yanai]{kurashige2011second}
Kurashige,~Y.; Yanai,~T. \emph{The Journal of chemical physics} \textbf{2011},
  \emph{135}, 094104\relax
\mciteBstWouldAddEndPuncttrue
\mciteSetBstMidEndSepPunct{\mcitedefaultmidpunct}
{\mcitedefaultendpunct}{\mcitedefaultseppunct}\relax
\EndOfBibitem
\bibitem[Roos(2003)]{roos2003ground}
Roos,~B.~O. \emph{Collection of Czechoslovak chemical communications}
  \textbf{2003}, \emph{68}, 265--274\relax
\mciteBstWouldAddEndPuncttrue
\mciteSetBstMidEndSepPunct{\mcitedefaultmidpunct}
{\mcitedefaultendpunct}{\mcitedefaultseppunct}\relax
\EndOfBibitem
\bibitem[Bondybey and English(1983)Bondybey, and
  English]{bondybey1983electronic}
Bondybey,~V.; English,~J. \emph{Chemical Physics Letters} \textbf{1983},
  \emph{94}, 443--447\relax
\mciteBstWouldAddEndPuncttrue
\mciteSetBstMidEndSepPunct{\mcitedefaultmidpunct}
{\mcitedefaultendpunct}{\mcitedefaultseppunct}\relax
\EndOfBibitem
\bibitem[Casey and Leopold(1993)Casey, and Leopold]{casey1993negative}
Casey,~S.~M.; Leopold,~D.~G. \emph{The Journal of Physical Chemistry}
  \textbf{1993}, \emph{97}, 816--830\relax
\mciteBstWouldAddEndPuncttrue
\mciteSetBstMidEndSepPunct{\mcitedefaultmidpunct}
{\mcitedefaultendpunct}{\mcitedefaultseppunct}\relax
\EndOfBibitem
\bibitem[Hilpert and Ruthardt(1989)Hilpert, and Ruthardt]{hilpert1989ber}
Hilpert,~K.; Ruthardt,~K. \emph{Ber. Bunsenges.} \textbf{1989}, \emph{93},
  1070--1078\relax
\mciteBstWouldAddEndPuncttrue
\mciteSetBstMidEndSepPunct{\mcitedefaultmidpunct}
{\mcitedefaultendpunct}{\mcitedefaultseppunct}\relax
\EndOfBibitem
\bibitem[Su \latin{et~al.}(1993)Su, Hales, and Armentrout]{su1993bond}
Su,~C.-X.; Hales,~D.~A.; Armentrout,~P. \emph{Chemical physics letters}
  \textbf{1993}, \emph{201}, 199--204\relax
\mciteBstWouldAddEndPuncttrue
\mciteSetBstMidEndSepPunct{\mcitedefaultmidpunct}
{\mcitedefaultendpunct}{\mcitedefaultseppunct}\relax
\EndOfBibitem
\end{mcitethebibliography}

\end{document}